\documentclass[aps,prl,twocolumn,showkeys,superscriptaddress]{revtex4}

\usepackage{amssymb}
\usepackage{graphicx}
\usepackage{xcolor}
\usepackage{amsmath}
\usepackage{amsfonts}
\usepackage{amssymb}
\usepackage{braket}
\usepackage{placeins} 

\begin{document}

\title{Thermoelectric transport in a correlated electron system on the surface of liquid helium}

\author{Ivan Kostylev}
\email[E-mail: ]{ivan.kostylev@oist.jp}
\author{A. A. Zadorozhko}
\author{M. Hatifi}
\affiliation{Quantum Dynamics Unit, Okinawa Institute of Science and Technology, Tancha 1919-1, Okinawa 904-0495, Japan}
\author{Denis Konstantinov}
\email[E-mail: ]{denis@oist.jp}
\affiliation{Quantum Dynamics Unit, Okinawa Institute of Science and Technology, Tancha 1919-1, Okinawa 904-0495, Japan}
\date{\today}

\begin{abstract}
We report on the direct observation of the thermoelectric transport in a nondegenerate correlated electron system formed on the surface of liquid helium. We find that the microwave-induced excitation of the vertical transitions of electrons between the surface-bound states leads to their lateral flow, which we were able to detect by employing a segmented electrode configuration. We show that this flow of electrons arises due to the Seebeck effect, thus our method provides a new tool to study thermoelectricity in a disorder-free correlated electron system. Our experimental results are in good agreement with the theoretical calculations based on kinetic equations, with proper account of fast electron-electron collisions. 
\end{abstract}

\maketitle
  
Recently there has been a renewed interest in the thermoelectric transport properties of materials~\cite{SnydNatMat2008}. Thermoelectric power generation, that is conversion of the thermal energy into electric power under a temperature gradient through the Seebeck effect, is believed to be one of the key technologies for sustainable energy. Driven by desire to improve efficiency of the energy-conversion process the interest has turned to systems other than conventional bulk solid-state semiconductors, such as low-dimensional mesoscopic structures~\cite{NaraCRPhy2016,HochNat2008,BoukNat2008,ShimPNAS2016}, graphene sheets and nanostructures~\cite{ZongACS2020}, as well as organic-based materials~\cite{RussNatMat2016}. The thermoelectric performance is a tradeoff between different transport parameters which intrinsically are determined by carrier scattering mechanisms in the system. Besides a large value of the Seebeck coefficient $S$, which relates the temperature gradient and the generated electromotive force, it is desired to have a large value of the electrical conductivity $\sigma$ and a low value of the thermal conductivity $\kappa$. This is summarized in the dimensionless figure of merit $ZT$ at a given temperature $T$ as $ZT=(S^2\sigma/\kappa)T$. The thermoelectric performance of modern solid-state systems is still poor, with typical $ZT\lesssim 2$~\cite{VenkNat2001,PoudSci2008,BiswNat2012}. The challenge for increasing $ZT$ arises from the coupling between the transport coefficients $S$, $\sigma$ and $\kappa$, and decoupling these parameters is nontrivial.  In particular, it was inferred that study of the thermoelectric transport in controllable model systems, such as cold atomic gases~\cite{BranSci2013,GrenRCPhys2016}, is important for fundamental understanding of the problem.   

Electrons trapped on the surface of liquid helium present an extremely clean and controllable two-dimensional (2D) nondegenerate Coulomb system~\cite{Monarkha-book}. The surface-bound  states of such electrons are formed due to an attractive image-charge potential and a repulsive barrier at the surface. Trapped electrons are free to move laterally along the surface and scatter only from helium vapor atoms, whose density is negligible at temperatures sufficiently below 1 K, or the surface capillary waves (ripplons). Thus, the system is free of static disorder typical for solid-state systems. Electronic transport, which has been extensively studied in this system, reveals the highest electron mobility known in any charge systems~\cite{Andrei-book}. Much less is known about the thermal and thermoelectric transport of surface electrons (SE) on helium. Recently it was demonstrated that the Seebeck effect can be  used to probe the temperature of SE in the vapor-atom scattering regime at 1.6 K~\cite{LyonPRL2018}. In this work, the electron-electron interaction was omitted from the discussion. However, theoretically it is expected that very fast inelastic electron-electron scattering, which is a signature of strongly-correlated electron liquid on liquid helium~\cite{ZipfPRL1976}, can lead to strong violation of the Wiedemann-Franz law and significant increase of the ratio $\sigma/\kappa$~\cite{PrinPRL2015,LucaPRB2018,LeePRR2020}. Thus, a careful study of thermoelectric transport in this controllable disorder-free system can potentially help us to explore different regimes of thermoelectric transport for energy-conversion optimization.

In this Letter, we report direct observation of the thermoelectric transport of electrons on the surface of liquid helium. We used resonant microwave excitation of the electron transitions between surface-bound states to heat SE locally with a high spatial resolution, while, by employing a segmented geometry of the electrodes coupled to SE, we were able to extract a signal corresponding to charge flow along the surface. We find that the net particle current is always directed from the hot place to the cold place, which is the signature of the Seebeck effect. The magnitude of the thermoelectric current, its nonlinear microwave-power dependence, as well as its dynamic response, are all in a good agreement with our calculations based on the kinetic equations. Interestingly, by taking into account extremely fast inelastic electron-electron collisions, the theoretically estimated figure of merit $ZT$ can reach values much larger than unity due to strong violation of the Weidemann-Franz law. We conclude that our experimental method demonstrates a new tool to study thermoelectricity in a disorder-free correlated charged particle system.      

\begin{figure} 
    \includegraphics[width=8.5cm]{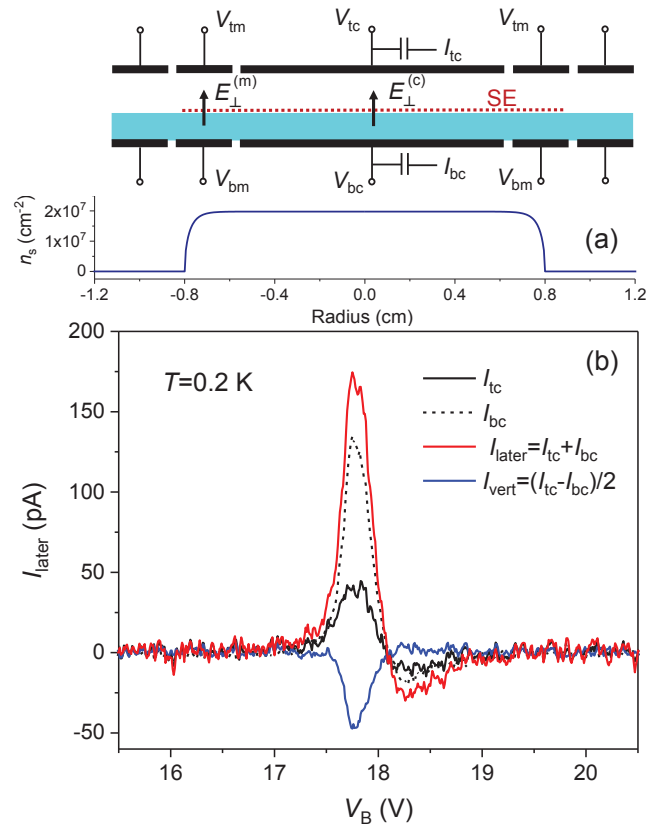}
    \caption{(color online) (a) Schematic view of the segmented electrode geometry used in the experiment and the equilibrium electron density profile calculated for $V_\textrm{bc}=V_\textrm{bm}=17.8$~V and all other electrodes grounded. (b) The image-current response obtained at $T=200$~mK using pulse-modulated ($f_m=400$~kHz) excitation of SE by 130-GHz radiation. The measured image currents $I_\mathrm{tc}$ and $I_\mathrm{bc}$ at the central segments of the capacitor's top and bottom plates, respectively, are plotted by solid (black) and dashed  (black) lines, respectively.}
    \label{fig:1}
\end{figure}

In our experimental setup  SE, of typical number density $n_s \approx 10^7$~cm$^{-2}$, are trapped on a free surface of liquid $^3$He placed between two circular electrodes and cooled below 1K inside a leak-tight experimental cell attached to the cold plate of a refrigerator. The electrodes, each of 24~mm diameter, form a parallel-plate capacitor with a gap $d =2.1$~mm between the top and bottom plates, and the liquid level is placed approximately midway between the plates. Electrons form a round pool of surface charge when a positive voltage is applied to the bottom plate. SE can be exposed to a microwave (MW) radiation at the frequency $\sim 100$~GHz transmitted from a room temperature source into the cell via a waveguide. The frequency of transitions between the surface-bound states of SE can be tuned in resonance with the applied radiation via the Stark shift by varying the electric field $E_\perp$ in the capacitor gap. As was recently demonstrated, the vertical displacement of electrons due to their excitation to higher surface states can be detected by measuring the change in the image charge induced by SE at the capacitor plates~\cite{KawaPRL2019}. In particular, a pulse-modulated MW excitation causes a modulated image current in the plates whose first harmonic can be measured by a conventional lock-in amplifier. It is clear that the image-charge currents induced in the top and bottom plates must have the same magnitude and opposite polarities, as was indeed confirmed in the experiment~\cite{KawaPRL2019}. In the experiments described here, a simple but important modification of this experimental setup was introduced. Particularly, each round plate of the capacitor was segmented into three concentric electrodes by two circular gaps of diameters 13 and 18~mm, each having a width of about 0.2~mm, as shown schematically in Fig.~\ref{fig:1}(a). By applying separate electrical potentials to different segments, the size of the electron pool, as well as the value of the tuning electric field $E_\perp$ at different parts of the system, can be varied. As an example, the calculated SE density profile when a positive potential is applied to the central and middle segments of the bottom electrode ($V_\mathrm{bc}=V_\mathrm{bm}=+17.8$~V), while all other segments are grounded, is shown in Fig.~\ref{fig:1}(a). Additionally, the image currents due to displacement of SE can be measured at each segment independently. As shown below, such a setup allowed us to reveal a strong lateral motion of SE due to pulse-modulated resonant MW excitation, in addition to the vertical displacement of SE due to occupation of the higher surface states.

Fig.~\ref{fig:1}(b) shows the image currents $I_\mathrm{tc}$ (solid black line) and $I_\mathrm{bc}$ (dashed black line) at the central segments of the top and bottom plates, respectively, for SE under pulse-modulated MW excitation at frequency $\omega/2\pi=130$~GHz for different values of a positive voltage $V_\textrm{b}$ $(=V_\mathrm{bc}=V_\mathrm{bm})$ applied simultaneously to the central and middle segments of the bottom plate. A nonzero current response around $V_\textrm{b}=17.8$~V indicates the MW-induced excitation of SE from the ground state to the first-excited surface state. The corresponding value of the tuning electric field $E_\perp=V/d\approx 85$~V/cm is in a reasonable agreement with the corresponding transition frequency for the Stark-shifted surface states~\cite{Monarkha-book}. Contrary to what is expected from the vertical displacement of electrons, the magnitude of measured currents $I_\mathrm{tc}$ and $I_\mathrm{bc}$ differs significantly. It is clear that at resonance, in addition to the vertical displacement of SE seen earlier~\cite{KawaPRL2019}, there is also a strong in-plane displacement of SE. Thanks to our geometry, it is straightforward to eliminate the contribution to measured currents from the vertical displacement of SE by adding the two measured currents together. The sum of two image currents, $I_\mathrm{later}=I_\mathrm{bc}+I_\mathrm{tc}$, is shown by a solid red line in Fig.~\ref{fig:1}. This signal corresponds to the depletion of the negative surface charge due to SE located between the central segments, therefore the lateral displacement of SE away from the center of the electron pool. We have confirmed that the sum of two currents measured at the middle segments of bottom and top plates, respectively, is nearly equal in magnitude and opposite in sign to $I_\mathrm{later}$, as should be expected from the particle conservation. Also, since the lateral displacement of SE must induce equal changes in the image charges at the top and bottom segments, providing the liquid level is at the middle between the capacitor plates, the signal due to the vertical displacement of SE induced, for example, at the top plate can be recovered by subtracting the measured currents, that is $I_\mathrm{vert}=(I_\mathrm{tc}-I_\mathrm{bc})/2$. This quantity is shown by the solid blue line in Fig.~\ref{fig:1}(b). As expected, this signal is similar to the one measured earlier~\cite{KawaPRL2019}.

\begin{figure} 
    \includegraphics[width=8.5cm]{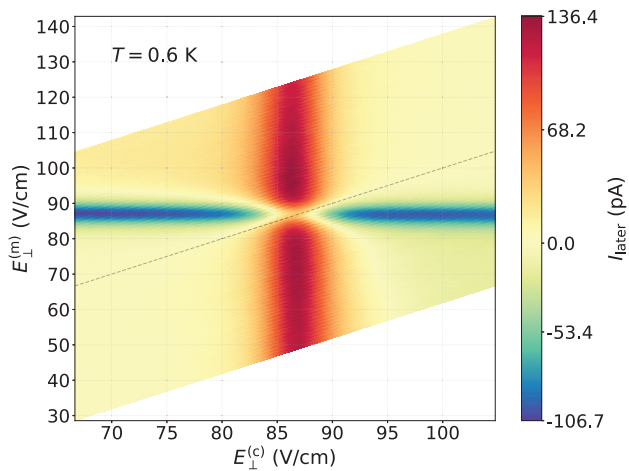}
    \caption{(color online) Color map of the measured lateral displacement current $I_\mathrm{later}$ versus $E_\perp^{(c)}$ and $E_\perp^{(m)}$ for SE under pulse-modulated ($f_m=200$~kHz) excitation by 130 GHz-radiation at $T=600$~mK. The dashed line marks $E_\perp^{(c)}=E_\perp^{(m)}$.}
    \label{fig:2}
\end{figure}

Using this procedure we can investigate, in details, dependence of the electron lateral displacement on various experimental parameters. Of particular interest is the displacement direction within the electron pool. For the data shown in Fig.~\ref{fig:1}(a), SE occupying areas between the central and middle segments are tuned in resonance simultaneously and $I_\mathrm{later}$ is positive. The segmented geometry of electrodes allows us to tune these two parts of the electron pool independently while retaining the same electrical potentials at the surface to keep the same equilibrium electron density profile. To accomplish this, we apply the following voltages: $V_\mathrm{bc}=V_1$, $V_\mathrm{tc}=0$, $V_\mathrm{bm}=V_1+V_2$, and $V_\mathrm{tm}=-V_2$, where $V_1$ and $V_2$ can be varied independently. It is clear that this configuration of potentials satisfies the above requirements. Particularly, the tuning electric fields for SE between the central and middle segments are given by $E_\perp^{(c)}=V_1/d$ and $E_\perp^{(m)}=(V_1+2V_2)/d$, respectively, as shown schematically in Fig.\ref{fig:1}(a). Fig.~\ref{fig:2} shows a color map of $I_\mathrm{later}$ versus $E_\perp^{(c)}$ and $E_\perp^{(m)}$ obtained for SE under pulse-modulated MW excitation as described above. Particular care was taken to ensure a uniform density profile for SE above the middle electrode. The lateral displacement current is positive (red color) when SE at the central part of the electron pool are tuned in resonance with the applied radiation. Contrarily, the current is negative (blue color) when SE in the peripheral part of the electron pool are tuned to the resonance. This corresponds to the lateral displacement of SE from the peripheral part of the electron pool towards the center. The data in Fig.~\ref{fig:2} unequivocally confirms that under irradiation SE are displaced from the part of the pool where they are resonantly excited towards the part where they are off resonance. In particular, this explains the change in the sign of $I_\mathrm{later}$ on the high-voltage side of the resonance peak in Fig.~\ref{fig:1}(b). Indeed, at such fields SE at the edge of the electron pool, which experience somewhat lower $E_\perp$-field than the rest of SE, become tuned to the resonance, while the rest of the pool is already detuned. Correspondingly, electrons displace from the resonantly-excited edge of the pool towards its center, which results in $I_\mathrm{later}<0$.    

It is well established that MW-induced excitation of the surface states of SE on liquid helium heats the electron system due to decay of the excited SE caused by elastic collisions with vapor atoms and ripplons~\cite{VoloJETP1981,KonsPRL2007}. Such collisions, in a typical time scale $10^{-8}$-$10^{-10}$~s in the temperature range 0.2-0.6~K considered here, provide an effective channel to transfer excitation energy absorbed by SE from the MW field to the electron thermal energy~\cite{KonsPRL2007}. The thermal equilibrium among the irradiated electrons, with an effective electron temperature $T_e>T$, is established on a time scale of the inelastic electron-electron collisions, which is in the order $10^{-10}$~s for electron densities considered here~\cite{ZipfPRL1976}. The electron temperature is determined by balance between the energy gain due to MW absorption and energy loss due to inelastic collisions of SE with scatterers. The latter is slow, in a time scale $10^{-6}$-$10^{-7}$~s for typical temperatures considered here~\cite{KawaPRL2021}. This allows overheating of the MW-excited SE to a few degrees Kelvin above the bath temperature $T$ for typical MW powers used in the experiment. Finally, the heat can propagate in the electron system through the electronic thermal conductance.  

The thermoelectric transport in SE must be described by kinetic equations taking into account elastic scattering of electrons from vapor atoms and rippons, as well as inelastic electron-electron collisions. According to Ref.~\cite{LeePRR2020} (see also an earlier work by Keyes~\cite{KeyeJPCS1958}), the Boltzmann equation in the relaxation-time approximation for a many-electron system can be presented as

\begin{equation}
\frac{\partial f}{\partial t} + \textbf{v} \nabla f +\dot{\textbf{p}} \nabla_\textbf{p} = -\frac{f-f_0}{\tau} - \frac{f-f_0^*}{\tau_{ee}}, 
\end{equation}

\noindent where $f=e^{(\mu-\epsilon)/\beta}$ is the distribution function for nondegenerate electrons ($\mu$ is the chemical potential, $\varepsilon$ is the electron total energy, and $\beta=(k_BT_e)^{-1}$), $f_0$ and $f_0^*$ is the equilibrium distribution in the laboratory and center-of-mass frames, respectively, $\tau$ is the electron momentum relaxation time due to elastic scattering from vapor atoms and ripplons, and $\tau_{ee}$ is the electron-electron scattering time. Following Ref.~\cite{LeePRR2020}, it is straightforward to obtain the thermoelectric transport coefficients $S$, $\sigma$ and $\kappa$. For the stationary state of the electron system, where no electric current flows, we obtain~\cite{SM}  

\begin{eqnarray}
&& \sigma=\frac{e^2 n_s \tau}{m_e}, \quad S=-\frac{k_B}{e} \left( 2-\mu\beta + \langle E_n \rangle\beta \right), \nonumber \\
&& \kappa=\frac{n_s\tilde{\tau}}{m_eT_e} \left( 2\beta^{-2} + \langle E_n^2 \rangle - \langle E_n \rangle ^2 \right).
\end{eqnarray}   

\noindent Here, $\tilde{\tau}=(\tau^{-1}+\tau_{ee}^{-1})^{-1}$, $E_n$ is the quantized energy of electron vertical motion, and $\langle .. \rangle$ denotes the thermal average. Note that, for the sake of simplicity, we consider momentum-independent scattering time $\tau$, which is valid for elastic scattering by vapor atoms dominating at $T>0.3$~K for SE on liquid $^3$He. Interestingly, the figure of merit $ZT$ obtained from the above equations can significantly exceed unity. Indeed, by using the estimates $S\approx -2(k_B/e)$ and $\kappa \approx (2k_BT_e n_s\tilde{\tau})/m_e$, we obtain $ZT\approx 2(\tau/\tilde{\tau})$, which varies between about 350 and 4 for the temperature range between 0.2 and 0.6 K. Such large values of $ZT$ are due to $\tau_{ee} << \tau$, which corresponds to strong violation of the Weidemann-Franz law~\cite{PrinPRL2015,LucaPRB2018,LeePRR2020}. 

Using the above expressions and azimuthal symmetry of our experimental setup, we can find the current response of the electron system to resonant heating by solving the coupled partial differential equations for electron temperature $T_e(r,t)$ and electron density $n_s(r,t)$ 

\begin{eqnarray}
&& \frac{\partial T_e}{\partial t} = W_\textrm{abs} - W_\textrm{loss} + \frac{1}{n_s r} \frac{\partial }{\partial r} \left( \kappa r \frac{\partial T_e}{\partial r}\right), \nonumber \\
&& \frac{\partial n_s}{\partial t} = \frac{\sigma}{C r} \frac{\partial}{\partial r}\left( r \frac{\partial n_s}{\partial r}\right) - \frac{\sigma S}{e r} \frac{\partial}{\partial r}\left( r \frac{\partial T_e}{\partial r}\right).  
\label{eq:coupl}  
\end{eqnarray} 

\noindent The first expression is the heat conduction equation, where $W_\textrm{abs}$ and $W_\textrm{loss}$ (divided by $k_B$) is the rate of energy absorption and energy loss per electron, respectively. The former depends on the value of MW electric field $E_\textrm{MW}$ inside the cell, while the latter is determined by the inelastic scattering of electrons. The second expression is the continuity equation, where we used relation for the electric current density $j_e = - \sigma\nabla \phi - \sigma S \nabla T_e$~\cite{SM}. The electric potential of SE $\phi$ is related to their density by the Poisson equation. Here, for the sake of simplicity we used a capacitance model for small variations of $\phi$ and $n_s$ around their equilibrium values, $\delta\phi = -e\delta n_s/C$, where $C=2\epsilon_0/d\approx 1.8\times 10^{-8}$~F/m$^2$~\cite{SM}.    

\begin{figure}
    \includegraphics[width=8.5cm]{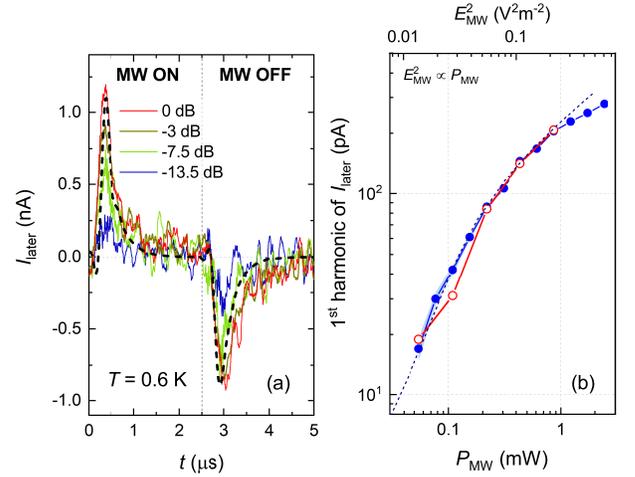}
    \caption{(color online) (a) Measured transient traces of $I_\textrm{later}(t)$ for SE under pulsed-modulated ($f_\textrm{m}=200$~kHz) 130-GHz radiation for different values of the MW-source power (indicated in dB with respect to the maximum power). The dashed black line is the current response calculated using Eqs.~(\ref{eq:coupl}) for $E_\textrm{MW}=0.42$~V/m. The small ripple effect at $t=0$ and $t=2.5~\mu$s on the calculated trace is due to the 2~MHz low-pass filtering. (b) The first harmonic of $I_\textrm{later}(t)$ measured by the lock-in amplifier (open red circles) and extracted from transient traces shown in panel (a) (closed blue circles) versus input MW power $P_\textrm{MW}$. Dashed line represents $I_\textrm{later}$ versus $E_\textrm{MW}^2$ extracted from the calculated transient traces~\cite{SM}.}
    \label{fig:3}
\end{figure}  

In order to make comparison with our calculations, we performed real-time measurements of $I_\textrm{later}(t)$ for SE tuned in resonance with the pulse-modulated MW radiation using a fast current preamplifier (FEMTO\textsuperscript{\textregistered} DHPCA-100). The transient current traces for different values of the output power $P_\textrm{MW}$ of our MW source are shown in Fig.~\ref{fig:3}(a). For the sake of comparison, a representative trace calculated using Eqs.~(\ref{eq:coupl}) for $E_\textrm{MW}=0.42$~V/m, with a 2~MHz low-pass filter applied to match the experimental conditions~\cite{SM}, is also shown by the dashed black line. Fig.~\ref{fig:3}(b) shows the power dependence of $I_\textrm{later}$ measured directly by the lock-in amplifier (open red circles), as well as the first harmonic extracted from the measured transient current traces (closed blue circles). As expected, we get essentially identical results. For the sake of comparison, the first harmonic extracted from the calculated traces for different values of $E_\textrm{MW}^2 \propto P_\textrm{MW}$ is plotted by the dashed line. Note that an exact relation between the output power $P_\textrm{MW}$ and the electric field $E_\textrm{MW}$ can not be accurately determined. For this reason, the range of $E_\textrm{MW}$ (the top axis) is adjusted to give the best match between the theory and experiment. The observed nonlinearity of the power dependence, which stems from the non-linear power dependence of $T_e$, is reproduced well by our calculations. We note that this peculiar nonlinearity allows us to rule out some other explanations of the observed lateral current, for example due to a photogalvanic effect~\cite{EntiJETP2013}, which would produce a linear power dependence.         

In conclusion, we described direct observation of the thermoelectric transport in a 2D electron liquid due to a nonuniform heating of the electron system by resonant MW radiation. Our results are in good agreement with the calculations based on kinetic equations. This work provides the opportunity to study thermoelectricity in a clean model system with controllable electron-electron interaction, which can affect electric and thermal properties of the system in a nontrivial way~\cite{PrinPRL2015,LucaPRB2018,LeePRR2020}. Our simple experimental method can be readily adjusted to different configurations by choosing an appropriate electrode geometry. In particular, an interesting problem is to study thermoelectricity of strongly-correlated SE in the presence of a periodic potential commensurate with the spacing between electrons~\cite{DimaEPL2013}. Such an experiment can be potentially realized in microchannel structures filled with superfluid helium~\cite{LinJLTP2018}, which is an interesting alternative to the proposed experiments with cold ion systems~\cite{DimaEPJ2019}.        

\medskip
\begin{acknowledgments}
This work was supported an internal grant from Okinawa Institute of Science and Technology (OIST) Graduate University.
\end{acknowledgments}


\begin{thebibliography}{99}

\bibitem{SnydNatMat2008} G. J. Snyder and E. S. Toberer, Nat. Mater. \textbf{7}, 105 (2008).

\bibitem{NaraCRPhy2016} V. Narayan, M. Pepper, and D. A. Ritchie, C. R. Physique \textbf{17}, 1123 (2016).

\bibitem{HochNat2008} A. I. Hochbaum, R. Chen, R. D. Delgado, W. Liang, E. C. Garnett, M. Najarian, A. Majumdar, and P. Yang, Nature \textbf{451}, 163 (2008).

\bibitem{BoukNat2008} A. I. Boukai, Y. Bunimovich, J. Tahir-Kheli, J.-K. Yu, W. A. Goddard III, and J. R. Heath, Nature \textbf{451}, 168 (2008).

\bibitem{ShimPNAS2016} S. Shimizu, M. S. Bahramy, T. Iizuka, S. Ono, K. Miwa, Y. Tokura, and Y. Iwasa, PNAS \textbf{113}, 6438 (2016).

\bibitem{ZongACS2020} P. Zong, J. Liang, P. Zhang, C. Wan, Y. Wang, and K. Koumoto, ACS Appl. Energy Mater. \textbf{3}, 2224 (20020).

\bibitem{RussNatMat2016} B. Russ, A. Glaudell, J. J. Urban, M. L. Chabinyc, and R. A. Segalman, Nat. Rev. Mater. \textbf{1}, 16050 (2016).

\bibitem{VenkNat2001} R. Venkatasubramanian, E. Siivola, T. Colpitts, and B. O'Quinn, Nature \textbf{413}, 597 (2001).

\bibitem{PoudSci2008} B. Poudel, Q. Hao, Y. Ma, Y. Lan, A. Minnich, B. Yu, X. Yan, D. Wang, A. Muto, D. Vashaee, X. Chen, J. Liu, M. S. Dresselhaus, G. Chen, Z. Ren, Science \textbf{320}, 634 (2008).

\bibitem{BiswNat2012} K. Biswas, J. He, I. D. Blum, C.-I Wu, T. P. Hogan, D. N. Seidman, V. P. Dravid, and M. G. Kanatzidis, Nature \textbf{489}, 414 (2012).

\bibitem{BranSci2013} J.-P. Brantut, C. Grenier, J. Meineke, D. Stadler, S. Krinner, C. Kollath, T. Esslinger, A. Georges, Science \textbf{342}, 713 (2013).

\bibitem{GrenRCPhys2016} C. Grenier, C. Kollath, and A. Georges, C. R. Physique \textbf{17}, 1161 (2016).

\bibitem{Monarkha-book} Yu. P. Monarkha and Kono, {\it Two-dimensional Coulomb liquids and solids} (Springer-Verlag, Berlin, 2004).

\bibitem{Andrei-book} {\it Two-dimensional electron systems on helium and other cryogenic substrates}, edited by E.~Y. Andrei (Kluwer Academics, Dordrecht, MA, 1997). 

\bibitem{LyonPRL2018} E.~I. Kleinbaum and S.~A. Lyon, Phys. Rev. Lett. \textbf{121}, 236801 (2018).

\bibitem{ZipfPRL1976} C. L. Zipfel, T. R. Brown, and C. C. Grimes, Phys. Rev. Lett. \textbf{37}, 1760 (1976). 

\bibitem{PrinPRL2015} A. Principi and G. Vignale, Phys. Rev. Lett. \textbf{115}, 056603 (2015).

\bibitem{LucaPRB2018} A. Lucas and S. Das Sarma, Phys. Rev. B \textbf{97}, 245128 (2018).

\bibitem{LeePRR2020} W.-R. Lee, A. M. Finkel'stein, K. Michaeli, and G. Schwiete, Phy. Rev. Research \textbf{2}, 013148 (2020). 

\bibitem{KawaPRL2019} E. Kawakami, A. Elarabi, and D. Konstantinov, Phys. Rev. Lett. \textbf{123}, 086801 (2019).

\bibitem{VoloJETP1981} A. P. Volodin and V. S. \'Edel'man, Sov. Phys. JETP \textbf{54}, 198 (1981).

\bibitem{KonsPRL2007} D. Konstantinov, H. Isshiki, Yu. P. Monarkha, H. Akimoto, K. Shirahama, and K. Kono, Phys. Rev. Lett. \textbf{98}. 235302 (2007).

\bibitem{KawaPRL2021} E. Kawakami, A. Elarabi, and D. Konstantinov, Phys. Rev. Lett. \textbf{126}, 106802 (2021).

\bibitem{KeyeJPCS1958} R. W. Keyes, J. Phys. Chem. Solids \textbf{6}, 1 (1958).

\bibitem{WileJLTP1987} L. Wilen and R. Giannetta, J. Low Temp. Phys. \textbf{72}, 353 (1987).

\bibitem{SM} See Supplementary Material at (URL will be inserted by the publisher) for derivation of thermoelectric transport coefficients and numerical solution of time-dependent heat-conduction and continuity equations.

\bibitem{EntiJETP2013} M. V. Entin and L. I. Magarill, JETP Letters \textbf{98}, 816 (2013). 

\bibitem{DimaEPL2013} O. V. Zhirov and D. L. Shepelyansky, EPL \textbf{103}, 68008 (2013).

\bibitem{LinJLTP2018} J.-Y. Lin, A. V. Smorodin, A. O. Badrutdinov, and D. Konstantinov, J. Low Temp. Phys. \textbf{195}, 289 (2018).

\bibitem{DimaEPJ2019} O. V. Zhirov, J. Lages, and D. L. Shepelyansky, Eur. Phys. J. D \textbf{73}, 149 (2019).

\end{thebibliography}
\end{document}


\preprint{12}

\title{Supplemental Material for "Thermoelectric transport in a two-dimensional electron liquid on the surface of liquid helium"}

\author{Ivan Kostylev}
\affiliation{Quantum Dynamics Unit, Okinawa Institute of Science and Technology, Tancha 1919-1, Okinawa 904-0495, Japan}
\author{A. A. Zodorozhko}
\affiliation{Quantum Dynamics Unit, Okinawa Institute of Science and Technology, Tancha 1919-1, Okinawa 904-0495, Japan}
\author{M. Hatifi}
\affiliation{Quantum Dynamics Unit, Okinawa Institute of Science and Technology, Tancha 1919-1, Okinawa 904-0495, Japan}
\author{Denis Konstantinov}
\affiliation{Quantum Dynamics Unit, Okinawa Institute of Science and Technology, Tancha 1919-1, Okinawa 904-0495, Japan}

\maketitle

\section {Derivation of thermoelectric transport coefficients}

We start with Eq.~(1) in the main text and use an expansion of the electron distribution function $f_0^*$ in the center-of-mass frame $f_0^*\approx f - (\partial f/\partial \xi_p)(m_e\textbf{v}\cdot\textbf{v}_\textrm{c.m.})$,  where $\xi_p=\epsilon - \mu = p^2/(2m_e)+E_n - \mu$ and 

\begin{equation}
\textbf{v}_\textrm{c.m.} = N_e^{-1} \sum\limits_{\textbf{r}}\sum\limits_{\textbf{p},n} \textbf{v} (f-f_0), 
\label{eqsm:vcm}
\end{equation}

\noindent is the drift velocity of the center of mass of the electron system ($N_e$ is the total number of SE). Using the above expansion, we obtain

\begin{equation}
\frac{\partial f}{\partial t} + \textbf{v}\nabla f + \dot{\textbf{p}}\nabla_\textbf{p} f = -\frac{f-f_0}{\tilde{\tau}} - \left( \frac{\partial f}{\partial \xi_p} \right) \frac{m_e \textbf{v}\cdot\textbf{v}_\textrm{c.m.}}{\tau_{ee}}.
\end{equation} 

To obtain the thermoelectric transport coefficients, we consider stationary case $\partial f/\partial t=0$ and assume that there are spatial gradients of both the electron density $n_s$ (therefore the chemical potential $\mu$) and the electron temperature $T_e$, thus

\begin{equation}
f=f_0 + e\tilde{\tau} \textbf{v}\frac{\partial f}{\partial \xi_p} \left( -\nabla \phi + \frac{\nabla\mu}{e} - \frac{m_e\textbf{v}_\textrm{c.m.}}{e\tau_{ee}}\right) - \tilde{\tau} \textbf{v} \frac{\partial f}{\partial \xi_p}(\mu-\epsilon)\frac{\nabla T_e}{T_e},
\label{eqsm:df}
\end{equation} 

\noindent where the electrical potential $\phi$ is related to the force on an electron $\dot{\textbf{p}}=e\nabla\phi$ ($e>0$ is the elementary charge). Using Eqs.~(\ref{eqsm:vcm}) and (\ref{eqsm:df}) we can find $\textbf{v}_\textrm{c.m.}$ and plug it back to (\ref{eqsm:df}), thus obtain

\begin{equation}
f=f_0 + e\textbf{v}\frac{\tau_{ee}}{(\tau_{ee}-\tilde{\tau})} \frac{\partial f}{\partial \xi_p} \left( -\nabla\phi + \frac{\nabla\mu}{e}\right) -\tilde{\tau} \textbf{v}\frac{\partial f}{\partial \xi_p} (\mu-\epsilon)\frac{\nabla T_e}{T_e} - \textbf{v} \frac{\tilde{\tau}^2}{(\tau_{ee}-\tilde{\tau})} \textbf{v}\frac{\partial f}{\partial \xi_p} (\mu-\langle E_n\rangle-2k_BT_e)\frac{\nabla T_e}{T_e},
\end{equation}

\noindent where $\langle E_n\rangle$ is the mean energy of quantized vertical motion. The electrical current density $\textbf{j}_e$ and thermal current density $\textbf{j}_T=\textbf{j}_E + (\mu/e)\textbf{j}_e$ can be calculated using the above expression for $f$ by the usual way, which results in

\begin{eqnarray}
&& \textbf{j}_e = K_{11} \left( -\nabla\phi + \frac{\nabla\mu}{e} \right) + K_{12} \left( -\frac{\nabla T_e}{T_e} \right), \nonumber \\
&& \textbf{j}_T = K_{21} \left( -\nabla\phi + \frac{\nabla\mu}{e} \right) + K_{22} \left( -\frac{\nabla T_e}{T_e} \right),   
\label{eqsm:curr}
\end{eqnarray}

\noindent where

\begin{eqnarray}
&& K_{11}=\frac{e^2 n_s \tau}{m_e},  \\
&& K_{12}=K_{21}=\frac{e n_s \tau}{m_e} (\mu - \langle E_n\rangle -2k_B T_e),  \\
&& K_{22}=\frac{n_s\tilde{\tau}}{m_e} \left[ 6(k_BT_e)^2 -4(\mu -\langle E_n\rangle)(k_BT_e) + \mu(\mu-2\langle E_n\rangle) + \langle E_n^2\rangle + \left( \frac{\tilde{\tau}}{\tau_{ee}-\tilde{\tau}} \right) (\mu - \langle E_n\rangle -2k_BT_e )^2 \right]. \nonumber 
\label{eqsm:K}
\end{eqnarray}

\noindent For the electron-on-helium system, we must satisfy condition of zero particle flow at the stationary state, that is $\textbf{j}_e=0$. Using definition of the thermoelectric transport coefficients, we obtain $\sigma=K_{11}$, $S=K_{12}/(T_eK_{11})$, and $\kappa=T_e^{-1}(K_{22}-K_{12}^2/K_{11})$, which coincides with Eq. (2) in the main text.  

\section{Numerical solution of stationary and time-dependent heat-conductance and continuity equations}

\begin{figure}
    \includegraphics[width=16cm]{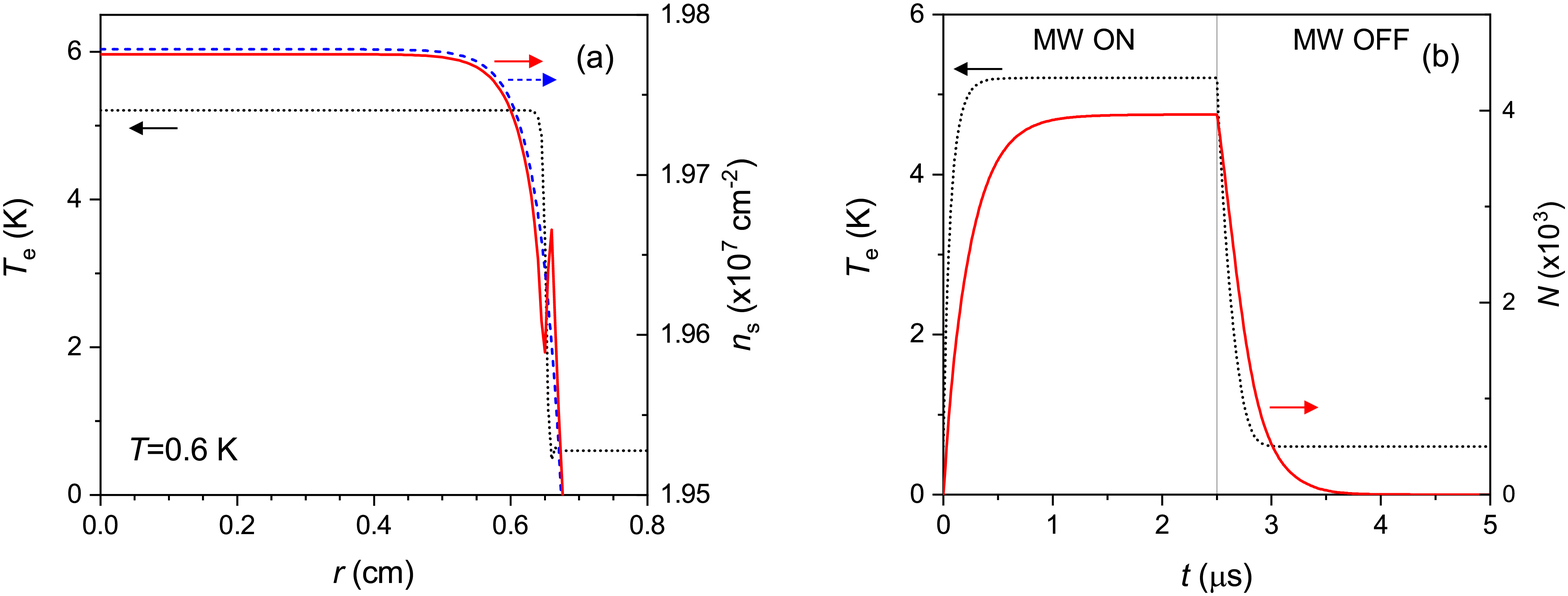}
    \caption{(color online) (a) Stationary spatial distribution of the electron temperature (dotted black line) and electron density (solid red line) calculated for SE at $T=0.6$~K when electrons at $r\leq 0.65$~cm are excited by MW radiation with $E_\textrm{MW}=0.42$~V/m, while the rest of SE are detuned from the resonance. The dashed blue line shows equilibrium density profile in the absence of heating. (b) Time evolution of the electron temperature at the center ($r=0$) of the SE system (dotted black line) and total number of electrons $N$ displaced from the excited region  at $r\leq 0.65$~cm (solid red line) calculated for pulse-modulated ($f_m=200$~kHz) MW radiation using the parallel-plate capacitance approximation, as described in the text.}
    \label{fig:S1}
\end{figure}  

The heat conduction equation, the first line in Eq.~(3) in the main text, expresses the balance between energy absorption, energy loss, and heat transfer by electrons in an azimuthally-symmetric system. The average (per electron) rate of energy gain due to the MW absorption at resonance can be expressed as $W_\textrm{abs}=(\rho_1 - \rho_2)\epsilon_\textrm{ph} (d_{12}E_\textrm{MW})^2/(2\hbar^2\gamma)$, where $\epsilon_\textrm{ph}$ is the MW photon energy, $d_{12}$ is the electric dipole transition moment, $\gamma$ is the half-width of the transition line, and $\rho_n$ is the fractional occupancy of the $n$-th energy level of the quantized vertical motion. The energy loss rate $W_\textrm{loss}$ is due to inelastic scattering of electrons by the vapor-atoms and ripplons and, in general, depends on the temperature of liquid helium $T$, electron temperature $T_e$, and occupancies $\rho_n$. The latter must be found self-consistently by solving the rate equations which include MW-induced transitions, as well as intersubband transitions due to elastic scattering of electrons from the vapor-atoms and ripplons. 

It is particularly convenient to do calculations for $T$ sufficiently above 0.3~K (the temperature of cross-over between vapor-atom and ripplon scattering regimes), where the scattering processes are dominated by the vapor atoms. In this case, the expressions for the corresponding scattering rates are significantly simpler than those for the ripplon scattering~\cite{KonsJPSJ2008}. Another advantage of high temperatures is that the intrinsic transition linewidth $\gamma$ due to vapor-atom scattering becomes comparable or larger than the inhomogeneous broadening due to a non-uniform electric field $E_\perp$ which Stark-shifts the energy levels of SE~\cite{AndoJPSJ1978}. This ensures a uniform excitation of the electron system tuned into the resonance. For this reason, we used $T=0.6$~K for the experimental data shown in Figs.~2 and 3 in the main text. 

Fig.~\ref{fig:S1} (a) shows an example of the stationary electron temperature profile calculated for SE at $T=0.6$~K when electrons between inner segments ($r\leq 0.65$~cm) of electrodes are excited by MW radiation with $E_\textrm{MW}=0.42$~V/m, while the rest of SE are detuned from the resonance. The total radius of the electron system is assumed to be $0.8$~cm (see Fig.~1(a) in the main text).  A very sharp temperature profile results from a low value of the thermal conductivity $\kappa$ determined by the short scattering time $\tilde{\tau}=(\tau^{-1}+\tau_{ee}^{-1})^{-1}$. Note that at $T=0.6$~K the values of $\tau$ and $\tau_{ee}$ are comparable and of the order $10^{-10}$~s. The electron temperature of the MW excited electrons is mostly determined by the balance between $W_\textrm{abs}$ and $W_\textrm{loss}$, which results in a "hot spot" at $r\leq 0.65$~cm with $T_e$ of several degrees Kelvin.

In order to find the corresponding charge displaced from the excited region of SE we numerically solve the coupled time-dependent equations for $T_e$ and $n_s$, see Eq.~(3) in the main text. The second line in this equation is the continuity equation $\partial n_s/\partial t = e^{-1} \nabla \textbf{j}_e$, where the electrical current $\textbf{j}_e = -\sigma \nabla (\phi - \mu/e) -\sigma S \nabla T_e$. It is convenient to use a local electrostatic approximation to relate the small variations in the electrical potential of SE and their density as $\delta\phi=-e\delta n_s/C$, with $C=2\varepsilon_0/d$, which follows from the parallel-plate capacitor model. This approximation is valid far from the edge of the electron system. Also, note that using the well-known relation $\nabla \mu = k_BT_e (\nabla n_s/n_s)$, it is easy to estimate that the term proportional to $\nabla \mu/e$ in the expression for $\textbf{j}_e$ can be safely neglected for typical experimental parameters. Finally, for the sake of simplicity we used an estimate for the Seebeck coefficient $S=-2(k_B/e)$, which is independent of $T_e$~\cite{LyonPRL2018}. The dotted black line in Fig.~\ref{fig:S1} (b) shows the time evolution of $T_e$ at the center ($r=0$) of the SE system in response to the pulse-modulated ($f_m=200$~kHz) excitation calculated by numerically solving the coupled time-dependent equation for $T_e$ and $n_s$ using the Partial Differential Equations (PDE) solver in Matlab. The solid red line shows the corresponding time-evolution of the total number of electrons $N$ obtained by integrating the density distribution $n_s(r,t)$ over the area of the central segment ($r\leq 0.65$~cm). The displaced charge of about $-4000e$ correspond to the mean value of current $Nef_m\approx 130$~pA, in a good agreement with the experimentally-observed values (see Fig.~3(b) in the main text). For the sake of comparison with the experimental results shown in Fig.~3(a) in the main text, the transient current response is obtained by differentiating the calculated displaced charge with respect to time and applying a low-pass digital filter ($f_c=2$~MHz) to account for the finite bandwidth of the current preamplifier (FEMTO\textsuperscript{\textregistered} DHPCA-100) used in the experiment. An exemplary result of such calculations is shown by the dashed black line in Fig.~3(a).  

In order to test the parallel-plate capacitance approximation, which significantly simplifies the above numerical calculations, we also calculated the stationary electron density profile $n_s(r)$ by solving the Poisson equation for the electrical potential taking into account condition for the stationary state $\nabla (\phi + ST_e)= 0$ (the term $\mu/e$ in the electrochemical potential is neglected, as described above). The Poisson equation was solved for our experimental cell geometry using the Green function method~\cite{WileJLTP1987}. The corresponding density profile calculated for the stationary electron-temperature profile shown in Fig.~\ref{fig:S1}(a) is shown by the solid red line in the same figure. For the sake of comparison, the equilibrium density profile in the absence of MW excitation is shown by the dashed blue line. The integrated displaced charge corresponds to about $-7500e$, which is in a reasonable agreement with the result obtained using the parallel-plate capacitance model.